\documentstyle[preprint,prl,aps,epsf]{revtex}

\newcommand{\bea}{\begin{eqnarray}} 
\newcommand{\ea}{\end{eqnarray}} 
\newcommand{\nn}{\nonumber\\}

\begin{document} 
 
\draft
\title{Dynamical Casimir effect at finite temperature}
\author{G\"unter Plunien, Ralf Sch\"utzhold, and Gerhard Soff}
\address{Institut f\"ur Theoretische Physik, Technische  Universit\"at
Dresden, \\
Mommsenstrasse 13, D-01062  Dresden, Germany}
\date{\today}
\maketitle

\begin{abstract} 
Thermal effects on the creation of particles under the influence of 
time-dependent boundary conditions are investigated.
The dominant temperature correction to the energy radiated by a 
moving mirror is derived by means of response theory. For a resonantly
vibrating cavity the thermal effect on the number of created photons
is obtained non-perturbatively. Finite temperatures can enhance the pure
vacuum effect by several orders of magnitude. 
The relevance of finite temperature 
effects for the experimental verification of the dynamical
Casimir effect is addressed.
\end{abstract}

PACS-numbers:  
42.50.Lc; 11.10.Wx; 11.10.Ef; 03.70.+k
  
\bigskip 

In 1948 H.~B.~G.~Casimir predicted an attractive 
force between two perfectly conducting, parallel plates
placed in the vacuum \cite{cas48}. 
A variety of fundamental and 
measurable consequences of quantum fluctuations under the influence 
of external conditions have been derived 
during the last decades (see e.g. \cite{revs} for reviews). With the
recent precision measurement of the Casimir force performed by
Mohideen {\em et al.} \cite{mar98} and by Lamoureaux \cite{lam97} conclusive
tests of Casimir's prediction are now available. They confirm the basic
concepts of quantum field theory in the presence of static external
constraints. 

One could expect that the success in measuring the static Casimir
force could also intensify experimental efforts in order to verify
a not less fundamental prediction, namely the {\em dynamical Casimir
effect}, i.e., the creation of particles out of the vacuum induced by
the interaction with dynamical external constraints. The phenomenon of 
vacuum radiation induced by moving mirrors \cite{moo70} initiated
intensive studies (see e.g. \cite{bae93} and references therein).
In particular the creation of photons in vibrating cavities seems to
be the most promising scenario for a possible experimental verification 
of motion-induced vacuum radiation 
(see e.g. \cite{law94,dod95,dak96,mag96,ljr96}). 

The thermodynamics of the static Casimir effect has been investigated 
intensively (see e.g. \cite{mar99} and references therein). 
In that context temperature effects 
are known to even dominate the pure vacuum effect (at $T=0$) and, in 
consequence, have to be taken into account when analysing the data in 
measurements of the static Casimir force. 
In contrast to this the dynamical Casimir effect at  finite temperature
so far has not been subject of research. 
It has been anticipated in a recent investigation by Lambrecht 
{\em et al.} \cite{ljr98} that temperature effects could play an 
important role for the generation of a photon pulse in a vibrating 
Fabry-P\'{e}rot cavity. 
However, realistic calculations of thermal effects on quantum radiation 
within the framework of quantum field theory 
of time-dependent systems at finite temperature are not yet available.  

Accordingly, it is our major intention to close this gap and to provide
a generalization of the Hamiltonian approach presented recently
in Refs. \cite{sps98a,sps98b}.
In this letter we focus the discussion 
predominantly on results obtained for the
thermal contribution to photon production in a resonantly vibrating cavity
as one of the most relevant configuration when aiming for experimental tests
of the dynamical Casimir effect \cite{dak96}. 
We like to address the question whether or not the effect
of quantum radiation might be covered by the thermal background and we will
examine the conditions under which it remains most significant even 
at large temperatures. Details of the derivations and further
applications of the formalism will be presented in a forthcoming publication
\cite{sps99}. 

We recall the canonical formalism in Ref. \cite{sps98a}, where
we considered a constrained, non-interacting, real, massless scalar field.
Similarly, the formalism also holds for bosonic quantum fields
interacting with classical external background fields \cite{sps98b}.
The boundary respectively the background may undergo small but arbitrary
dynamical changes resulting in an additional interaction Hamiltonian
$\hat H_{{\rm I}}$ which is assumed to be switched on and off at 
asymptotic times $t \rightarrow -\infty$ and $t \rightarrow +\infty$,
respectively.
For the boundary being initially at rest the (closed) system consisting 
of the scalar field enclosed by the boundary is assumed to be at
thermal equilibrium described by a statistical operator
$\hat\rho (t \rightarrow -\infty) = \hat\rho_0$. 
The existence of a total Hamiltonian $\hat H = \hat H_0 + \hat H_{{\rm I}}$
describing the evolution of the system implies that any kind of
backreaction of the quantum field upon the dynamics of the external 
constraints or any relaxation processes will be neglected. 
In addition, there should be no measurements on the quantum systems 
during the dynamical phase. We adopt the interaction representation. 
When the boundary or the background field experiences dynamical changes 
the system will no longer remain at thermal equilibrium. 
The time-evolution of the statistical
operator $\hat\rho$ is governed by the time-dependent interaction
Hamiltonian $\hat H_{{\rm I}}$ resulting in the von Neumann equation
(units where $\hbar = c = k_{\rm B} = 1$ are used throughout):
\bea
i\,\frac{d\hat\rho}{dt} = \left[\hat H_{{\rm I}},\hat\rho \right]\quad ,
\label{eq:1}
\ea
together with the initial condition $\hat\rho (t \rightarrow -\infty) = 
\hat\rho_0$. Here we disregard an explicit time-dependence
$(\partial \hat \rho /\partial t)_{{\rm exp.}}$ that could account for 
relaxation or backreaction processes.
Accordingly, this equation of motion can be integrated formally with the
aid of the time-evolution operator
\bea
\hat U (t',t)=
{\cal T}\left[\exp\left(-i\int_{t}^{t'} dt_1\,
\hat H_{{\rm I}}(t_1)\right)\right] 
\quad ,
\label{eq:2}
\ea
where ${\cal T}$ denotes time-ordering. 
The thermal expectation value at asymptotic times $t\rightarrow\infty$
of any relevant observable $\hat A$ is determined by
\bea
\langle\hat A\,\rangle
&=&{\rm Tr}\left\{\hat A\,\hat\rho(t\rightarrow\infty)\right\}
={\rm Tr}\left\{\hat A\, \hat U \, 
\hat\rho_0\,\hat U^+ 
\right\} \quad .
\label{eq:3}
\ea
The trace ${\rm Tr}\{\cdots \}$ involved is taken most conveniently over
the Fock space of the unperturbed Hamiltonian $\hat H_0$ refering to
the initial ensemble (with ${\rm Tr}\{\hat\rho_0 \} = 1$). 
The microscopic entropy 
$S = - {\rm Tr}\left\{\hat\rho \ln \hat\rho\right\}
= - {\rm Tr}\left\{\hat\rho_0 \ln \hat\rho_0\right\} = S_0$ 
remains constant in time. 
A more detailed examination of the thermodynamical aspects involved 
(e.g. effective entropy) will be given in \cite{sps99}. 
Eqs. (\ref{eq:1}) -- (\ref{eq:3}) generalize the canonical approach to
quantum radiation of Ref. \cite{sps98a,sps98b} to account for 
finite-temperature effects as well. 
In order to explore the influence of finite temperatures on the dynamical 
Casimir effect we have to investigate the thermal expectation value of
the number operator $\hat N_\lambda = \hat a^+_\lambda \hat a_\lambda$ 
of particles with frequency $\Omega^0_\lambda$. 
Note, that the proper definition of (quasi-) particle creation- 
(annihilation-) operators  $\hat a^+_\lambda$ ($\hat a_\lambda$) is 
provided with respect to the ground state $|0\rangle$
of the unperturbed Hamiltonian $\hat H_0$, which becomes diagonal, i.e., 
$\hat H_0 = \Omega^0_\lambda\,(\hat N_\lambda + 1/2)$. 
The total radiated energy associated with the dynamical Casimir-effect
can be deduced from the expectation value Eq. (\ref{eq:3})
of the normal-ordered unperturbed Hamiltonian 
${\rm Tr}\{:\!\hat H_0\!: \hat\rho\}$. 

In the following we consider the generic case of a constrained massless 
scalar or vector field at 
finite temperature $T=1/\beta$ described initially ($t\rightarrow -\infty$) 
by  the statistical operator (canonical ensemble)
\bea
\hat\rho_0=\exp\left(-\beta\hat H_0\right)/Z_0 \quad .
\label{eq:4}
\ea
The closed system is assumed to be initially at thermal equilibrium. 
The system leaves the thermal equilibrium when undergoing 
some dynamical changes.
Eq. (\ref{eq:3}) allows for a systematic perturbative approach in those
cases, for which a closed expression for the time-evolution operator is not
available. Especially for the thermal expectation value of the spectral
number density $\langle \hat N_\lambda\rangle$ one obtains up to quadratic
response  neglecting corrections of order ${\cal O} (\hat H_{{\rm I}}^3)$
\bea
\langle\hat N_\lambda\rangle
&=&
{\rm Tr}\left\{\hat N_\lambda\hat\rho_0\right\}
+
{\rm Tr}\left\{
\hat N_\lambda\,
\left[\int dt\,\hat H_{{\rm I}}(t),
\hat\rho_0\right]
\int dt\,\hat H_{{\rm I}}(t) 
\right\}
\nn
&=&
\langle\hat N_\lambda\rangle_0
+
\Delta N_\lambda\quad .
\label{eq:5}
\ea
The first term $\langle\hat N_\lambda\rangle_0 = 
{\rm Tr} \{\hat N_\lambda\,\hat\rho_0\}$ 
represents the usual thermal 
Bose-distribution function while the second term $\Delta N_\lambda$
denotes the motion-induced change of the number of particles (e.g. photons) 
with frequency $\Omega^0_\lambda$ at a given temperature $T$. The total
radiated energy, i.e. the change of the internal energy  
associated with quantum radiation is determined by
$\Delta E = \Omega^0_\lambda \Delta N_\lambda$ and can be deduced 
according to
$E = \langle :\!\hat H_0\!:\rangle = 
\langle :\!\hat H_0\!:\rangle_0 + \Delta E$. 

Let us now turn to the generic particle-creation processes.
In the case of constrained bosonic
quantum fields satisfying linear equations of motion the general 
form of the self-adjoint perturbation Hamiltonian induced by the
interaction with the external conditions assumes 
the rather general form (sum convention)
\bea
\int dt\,\hat H_{{\rm I}}(t)
=
\frac{1}{2}
\left(
{\cal S}_{\mu\nu}\hat a_\mu^+\hat a_\nu^+
+
{\cal S}_{\mu\nu}^*\hat a_\mu\hat a_\nu
\right)
+
{\cal U}_{\mu\nu}\hat a_\mu^+\hat a_\nu
+
{\cal C}
\label{eq:6}
\ea
with ${\cal S}_{\mu\nu}={\cal S}_{\nu\mu}$ 
and ${\cal U}_{\mu\nu}={\cal U}_{\nu\mu}^*$.
The constant ${\cal C}$ gives rise to a pure phase factor 
and thus drops out in
any expectation value (\ref{eq:3}). The ${\cal S}$-term may be interpreted 
as a generator of a multi-mode squeezing operator 
and the ${\cal U}$-term may be envisaged as a hopping operator.
Evaluating the quadratic response according to Eq. (\ref{eq:5}) 
the change of the number of particles yields
\bea
\Delta N_\lambda
&=&
\left|{\cal S}_{\lambda\rho}\right|^2
\left(1+\langle\hat N_\rho\rangle_0+\langle\hat N_\lambda\rangle_0\right)
\nonumber \\
&& +
\left|{\cal U}_{\lambda\rho}\right|^2
\left(\langle\hat N_\rho\rangle_0-\langle\hat N_\lambda\rangle_0\right)\quad .
\label{eq:7}
\ea
Note, that the $\cal S$-term contains the pure vacuum contribution 
$\Delta N_\lambda ({\rm vac}) = \left|{\cal S}_{\lambda\rho}\right|^2$.
One observes that only the ${\cal S}$-term contributes to the total 
number of created particles (sum over $\lambda$),
while the $\cal U$-term does not increase the total number of particles
(similar terms appear in master-equations). 
However, it modifies the configurations of
occupied particle states within the ensemble and thus 
also increases the total energy.

For trembling cavities the perturbation Hamiltonian 
(cf. Ref. \cite{sps98a} for details)
\bea
\hat H_{{\rm I}}(t) = \hat q_\mu^2(t)\, \Delta \Omega_\mu^2(t)/2
+ 
\hat q_\mu\,\hat p_\nu\,{\cal M}_{\mu\nu}(t)
\label{hint}
\ea
appears as a sum of a squeezing (first term) and a velocity term 
(second term).
The squeezing term involves the deviations $\Delta \Omega_\mu^2$ of 
the eigenmode spectrum 
induced by a change of the shape of the cavity, while the
velocity contribution involves the (antisymmetric) intermode couplings
${\cal M}_{\mu\nu}$ arising from the motion of the boundary.

In scenarios, where the velocity effect is supposed to be negligible 
the matrix ${\cal S}$ becomes diagonal
and the $\cal U$-term does not contribute at all. In view of 
Eq. (\ref{eq:7}) we identify 
the corresponding particle-production rate as the product of the pure 
vacuum-squeezing effect 
$\Delta N_\lambda^{S}({\rm vac})= \left|{\cal S}_{\lambda\lambda}\right|^2$ 
times a thermal distribution factor, i.e.
$\Delta N_\lambda^S = \Delta N_\lambda^{S}({\rm vac})\,
\left(1+2\langle\hat N_\lambda\rangle_0\right)$. It gives rise to a linear 
dependence on $T$ in the high-temperature limit.  

The pure velocity effect may be illustrated most simply by considering
a single moving mirror in 1+1 dimensions. For a single mirror placed 
at a position $\eta(t)$ the discrete eigenmodes $\Omega^0_\lambda$ 
have to be replaced by the continuous variable $k$. The total radiated
energy is derived as
\bea
E = 
\frac{1}{12\pi}\int dt\,\ddot\eta^2(t)+
\frac{\pi}{3}T^2\int dt\,\dot\eta^2(t) \quad ,
\label{eq:8}
\ea
which generalizes the zero-temperature result (first term) 
obtained by Fulling and Davis and by Ford and Vilenkin \cite{moo70}.
The ratio of the finite temperature correction 
to the radiated energy and the pure vacuum contribution
turns out to be of the order ${\cal O}(T^2 \tau^2)$, where $\tau$ denotes 
a characteristic time scale of the underlying dynamics. 

Let us now investigate the finite-temperature effects on the
dynamical Casimir-effect in a resonantly vibrating cavity. 
In order to allow for an experimental verification
the number of motion-induced created particles should be as large as possible.
One way to achieve this goal is to utilize the phenomenon
of parametric resonance, which occurs in the case of harmonically
time-dependent perturbations characterized by some frequency $\omega$. 
Obtaining large numbers may indicate that one has left 
the region, where second-order perturbation theory does apply. 

In the case of oscillating disturbances, however, it is possible to evaluate 
the time-evolution operator to all orders of $\hat H_{\rm I}$ 
analytically employing yet another approximation, the so-called
rotating wave approximation (see e.g. \cite{law94}). 
Let us assume that the explicit time-dependence of the 
perturbation Hamiltonian possesses an oscillatory behaviour like 
$ \varepsilon \sin(2\omega t)$ 
during a sufficiently long period of time {\sf T}, such that the conditions  
$\omega{\sf T}\gg 1$, $\varepsilon\ll 1$ and 
$\varepsilon\,\omega{\sf T}={\cal O}(1)$ hold. The main 
consequence of the rotating wave approximation consists in the fact 
that it allows for the derivation of a time-independent, effective
Hamiltonian $\hat H_\omega^{\rm eff} $ after performing the integration 
over time according to 
Eq. (\ref{eq:6}):
\bea
\int^{\sf T}_0 dt\,\hat H_{\rm I}\approx{\sf T}\,\hat H_\omega^{\rm eff}
\quad .
\label{eq:9}
\ea
This effective Hamiltonian accounts essentially for the dominant
contribution arising from the frequency $\omega$. 
For the evolution operator Eq. (\ref{eq:2}) this approximation implies
the neglection of all time-ordering effects
(all oscillating terms involving commutators average out). 

For a vibrating cavity the effective Hamiltonian $\hat H_\omega^{\rm eff}$ 
is easily calculated from the interaction operator (\ref{hint}). 
Assuming a harmonic time-dependence proportional to
$\varepsilon \sin (2\omega t)$ 
or $\varepsilon \cos (2\omega t)$ for both, the squeezing 
($\Delta\Omega_\mu^2$) and the velocity terms (${\cal M}_{\mu\nu}$)
only those terms will survive, which match the resonance conditions.
For the squeezing term the resonance condition reads 
$\Omega^0_\lambda=\omega$ and for the velocity term 
$|\Omega^0_\mu\pm\Omega^0_\nu|=2\omega$, respectively. Accordingly,  
the squeezing effect creates always particles with the
frequency $\omega$ provided this cavity mode does exist.
We restrict our further consideration to the 
situation, where the oscillation frequency $\omega$ corresponds to the 
lowest cavity mode 
$\omega ={\rm min}\left\{\Omega^0_\lambda\right\}=\Omega^0_1$. 
The fundamental resonance frequency is determined by
the characteristic size $L$ of the cavity, e.g.  $\Omega^0_1 = \sqrt{3}\pi/L$
for a cubic cavity.
For the lowest mode the resonance condition for  squeezing 
$\Omega^0_\lambda=\Omega^0_1$ is satisfied automatically.
Whether the resonance condition for the velocity effect
$|\Omega^0_\mu \pm \Omega^0_\nu|=2\Omega^0_1$ can be satisfied or not depends on the
spectrum of the particular cavity under consideration. 
For a one-dimensional cavity the eigenvalues $\Omega^0_\lambda$ are
proportional to integers and thus it can be satisfied
leading to an additional velocity contribution. For most cases of 
higher-dimensional cavities, e.g. a cubic one, 
this condition cannot be fulfilled. 
Thus the velocity effect does not contribute  
within the rotating wave approximation (cf. Ref. \cite{dak96}).

In consequence, only the squeezing term contributes. The effective  
Hamiltonian can be derived immediately for the lowest cavity mode.
From the contributing $\Delta \Omega_\lambda$-terms one obtains
$\hat H_1^{\rm eff} = i\Omega^0_1\varepsilon /4
\left[(\hat a_1^\dagger )^2 - (\hat a_1)^2\right]$
respectively for the time-evolution operator
\bea
\hat U({\sf T}, 0)  
\approx
\exp\left\{\frac{\Omega^0_1\varepsilon{\sf T}}{4}
\left[(\hat a_1^\dagger )^2-
(\hat a_1)^2\right]\right\}=\hat S_1
\quad .
\label{eq:13}
\ea
Note, that within this approximation $\hat U$ coincides 
with a squeezing operator $\hat S_1$ for the lowest mode $\lambda=1$.
This confirms the notion of the $\Delta \Omega^2_\mu$-terms in (\ref{hint})
as squeezing contribution.
After having derived a closed expression for the time-evolution operator 
this enables us to calculate the expectation value for the number operator
to all orders in $\hat H^{\rm eff}_1$
\bea
\label{quetsch}
\left<\hat N_\lambda\,\right>
\approx 
\left\langle\hat N_\lambda\right\rangle_0 +
\delta_{\lambda1}
\sinh^2\left(\frac{\varepsilon\Omega^0_1{\sf T}}{2}\right)
\left(1+2\left\langle\hat N_1\right\rangle_0\right)
\quad .
\label{eq:14}
\ea
This non-perturbative result implies that at finite temperature  
the number of photons $\Delta N_1$ created resonantly in the lowest cavity mode 
increases exponentially. The vacuum creation rate 
$\Delta N_1^{\rm S}({\rm vac}) = \sinh^2(\varepsilon\Omega^0_1{\sf T}/2)$ 
(see Ref. \cite{dak96}) gets enhanced by a thermal distribution factor. 

\begin{figure}[t]
\centerline{\mbox{\epsfxsize=15cm \epsffile{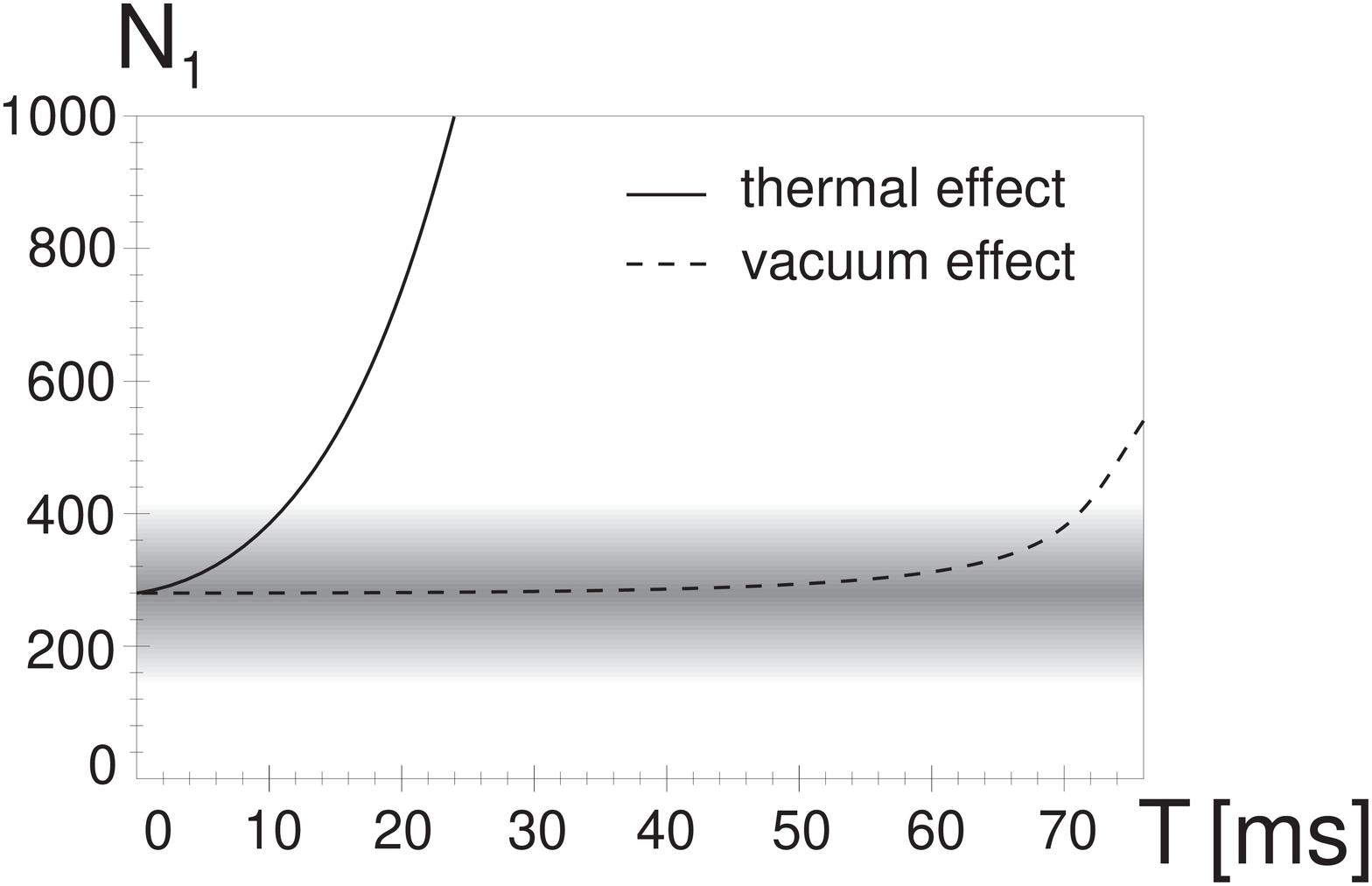}}}
\caption{ Number of photons $N_1$ produced via the dynamical Casimir effect
inside a resonantly vibrating cavity for a fixed value of the 
squeezing parameter $\varepsilon \omega /2 $ as a function of the
time duration {\sf T}. 
The fundamental frequency is choosen to be $\omega = 146$ GHz 
corresponding to a typical size of the cavity of $1$ cm.
For the dimensionless amplitude a value of $\varepsilon = 6\cdot 10^{-10}$ 
has been assumed. 
A finite temperature ($T = 290$ K) can enhance the 
effect (solid line) by several orders of magnitudes
compared with the pure vacuum effect at $T=0$ (dashed line).
The thermal background is indicated by the shadowed area
(half of the thermal variance $\sigma_0$ of the photon number $N_1$).}
\label{fig1}
\end{figure}

\bigskip
Equation (\ref{eq:14}) represents one essential result of our investigations.
In order to indicate its experimental relevance one may
specify the characteristic parameters. In Fig. \ref{fig1} the total number of
photons in the fundamental mode $\omega = \Omega^0_1$ inside of a cavity
of a characteristic size $L$ is depicted as a function of the time-duration
${\sf T}$ of the vibration of its wall. The contribution $\Delta N_1$
of additional, motion-induced photon creation 
at $T=0$, i.e. the pure vacuum effect
(dashed curve) is compared with the corresponding contribution at a 
temperature $T$ (solid curve). The shadowed area indicates the number of 
photons $\langle \hat N_1\rangle_0$ present at the temperature $T$
before the system will undergo dynamical changes together with its thermal
variance 
$\sigma_0(N_\lambda) = \sqrt{\langle\hat N_\lambda^2\rangle_0-
\langle\hat N_\lambda\,\rangle^2_0} \approx \langle\hat N_\lambda\,\rangle_0$
for $\langle\hat N_\lambda\,\rangle_0 \gg 1$. The latter reflects the 
uncertainty when measuring the number of photons at a given temperature $T$.
For room temperature $T\approx 290$ K, which corresponds to thermal 
wave lengths of about $50$ $\mu{\rm m}$ and considering a cavity of a typical
size $L\approx 1$ cm one obtains a thermal factor 
$(1+2\langle\hat N_1\,\rangle_0)$ of order $10^3$. 
As a consequence after the
vibration time ${\sf T}$ the number of photons $\Delta N_1$ created by the
dynamical Casimir effect at the given temperature will be three orders of
magnitude larger in comparison with the pure vacuum effect at $T=0$.  
At finite temperatures the dynamical Casimir effect 
should become observable even after shorter vibration times ${\sf T}$
as one would expect from looking at the pure vacuum effect.
This strong enhancement of the dynamical Casimir effect occuring at finite
temperatures could be exploited in  experiments to verify the
phenomenon of quantum radiation as long as backreaction processes can be 
neglected. Of course, one has to ensure conditions that will lead to
a significant vacuum effect as well. This implies that the argument of
the hyperbolic sine function in Eq. (\ref{eq:13}), i.e. the squeezing 
parameter $\varepsilon\omega {\sf T}/2$ should be at least of order $1$.
An estimate of the maximal value of the dimensionless amplitude
of the resonance wall vibration $\varepsilon_{{\rm max}} < 10^{-8}$ 
is given in Ref. \cite{dak96}. 
It still remains as a challenge whether or not the requirement
$\varepsilon\omega {\sf T}/2 \approx 1$ could be
achieved in a realistic experiment. 
We conclude: 
Provided an experimental device for generating a considerable vacuum
contribution becomes feasible, then we predict a strong enhancement 
of the dynamical Casimir effect at finite temperature.
From the theoretical point of view there is no definite
need to perform an even more involved experiment at low
temperatures.

\bigskip
G.~P. acknowledges enlightening discussions with Korn\'{e}l Sailer 
and Istv\'{a}n Lovas during his stay at the Department of Theoretical 
Physics at the Kossuth Lajos University in Debrecen, Hungary.
This visit was supported financially by the M\"OB and the DAAD.
Financial support from DFG, BMBF and GSI is also gratefully acknowledged.

\end{document}